\documentclass[aps,preprint]{revtex4}%
\usepackage{amsfonts}
\usepackage{amsmath}
\usepackage{amssymb}
\usepackage{graphicx}%
\begin{document}
\preprint{ }
\title{Pressure inequalities for nuclear and neutron matter\\}
\author{Dean Lee}
\affiliation{Department of Physics, North Carolina State University, Raleigh, NC 27695, USA}

\begin{abstract}
We prove several inequalities using lowest-order effective field theory for
nucleons which give an upper bound on the pressure of asymmetric nuclear
matter and neutron matter. \ We prove two types of inequalities, one based on
convexity and another derived from shifting an auxiliary field.

\end{abstract}
\maketitle

\section{Introduction}

In the effective field theory description of low-energy nuclear matter,
nucleons are treated as point particles rather than composite objects. \ While
much of the work in the community has focused on few-body systems, there has
also been recent interest in lattice simulations of bulk nuclear matter using
effective field theory
\cite{Muller:1999cp,Chen:2003vy,Abe:2003fz,Lee:2004si,Wingate:2004wm,Lee:2004qd}%
. \ In parallel with this computational effort, effective field theory was
also recently used to prove inequalities for the correlation function of
two-nucleon operators in low-energy symmetric nuclear matter \cite{Lee:2004ze}%
. \ It was shown that the $S=1$, $I=0$ channel must have the lowest energy and
longest correlation length in the two-nucleon sector. \ These results were
shown to be valid at nonzero density and temperature and could be checked in
effective field theory lattice simulations. \ The proof relied on positivity
of the Euclidean functional integral measure and is similar in spirit to
several quantum chromodynamics (QCD) inequalities proved using quark-gluon
degrees of freedom
\cite{Weingarten:1983uj,Vafa:1983tf,Vafa:1984xg,Vafa:1984xh,Witten:1983ut,Nussinov:1983vh,Nussinov:1999sx,Nussinov:2003uj,Cohen:2003ut}%
.

In this work we prove several new inequalities using effective field theory
which give an upper bound on the pressure of asymmetric nuclear matter and
neutron matter. \ We prove two types of inequalities, one based on convexity
and one derived from shifting an auxiliary Hubbard-Stratonovich field. \ We
consider two general types of systems, one with two fermion species and an
$SU(2)$ symmetry and another with four fermion species and an $SU(2)\times
SU(2)$ symmetry. \ The results we prove are quite general. \ In addition to
nuclear and neutron matter, our inequalities apply to systems of cold, dilute
gases of fermionic atoms
\cite{O'Hara:2002,Gupta:2002,Regal:2003,Bourdel:2003,Gehm:2003} which can be
described by the same lowest-order effective field theory.

\section{Lower\ bound}

Before deriving pressure upper bounds, we first state a general lower bound
for the pressure. \ The result is simple and perhaps obvious, but the
derivation is useful to help set our notation. \ Consider any system in
thermodynamic equilibrium that is invariant under a symmetry group $S$. \ Let
$\mu$ be a symmetric chemical potential which preserves the group $S$. $\ $Let
$\mu_{3}$ be an asymmetric chemical potential which breaks $S$ and flips sign
$\mu_{3}\rightarrow-\mu_{3}$ under some element of $S$. This means that the
pressure $P$ is an even function of $\mu_{3}$.

Our condition of thermodynamic equilibrium requires that the system\ is stable
and not further separating into regions with more widely different values of
$\mu_{3}$. $\ $This implies the convexity condition,%
\begin{equation}
\tfrac{\partial^{2}P(\mu,\mu_{3})}{\partial\mu_{3}^{2}}\geq0.
\end{equation}
Combining this with the fact that $P$ is even in $\mu_{3}$, we derive the
lower bound%
\begin{equation}
P(\mu,\mu_{3})\geq P(\mu,0)\text{.} \label{lower bound}%
\end{equation}
This lower bound holds for all the systems we consider here.

\section{Two fermion states - $SU(2)$}

We consider an effective theory with two species of interacting fermion fields
and an $SU(2)$ symmetry. \ Let $n$ be a doublet of fermion fields which we can
regard as neutron spin states,%
\begin{equation}
n=\left[
\begin{array}
[c]{c}%
\uparrow\\
\downarrow
\end{array}
\right]  .
\end{equation}
We can write the lowest-order Lagrange density in Euclidean space in two
equivalent forms,%

\begin{equation}
\mathcal{L}_{E}=-\bar{n}[\partial_{4}-\tfrac{\vec{\nabla}^{2}}{2m_{N}}%
+(m_{N}^{0}-\mu-\mu_{3}\sigma_{3})]n-\tfrac{1}{2}C\bar{n}n\bar{n}n,
\label{first neutron}%
\end{equation}
and%
\begin{equation}
\mathcal{L}_{E}=-\bar{n}[\partial_{4}-\tfrac{\vec{\nabla}^{2}}{2m_{N}}%
+(m_{N}^{0\prime}-\mu-\mu_{3}\sigma_{3})]n-\tfrac{1}{2}C^{\prime}\bar{n}%
\vec{\sigma}n\cdot\bar{n}\vec{\sigma}n, \label{second neutron}%
\end{equation}
where%
\begin{equation}
C^{\prime}=-\tfrac{1}{3}C.
\end{equation}
We use $\vec{\sigma}$ to represent Pauli matrices acting in spin space.
\ $\mu$ is the symmetric chemical potential while $\mu_{3}$ is the asymmetric
chemical potential. \ We assume the interaction is attractive so that
\begin{equation}
C<0\text{, }C^{\prime}>0\text{.}%
\end{equation}

\subsection{Two-body operator coefficients}

\begin{figure}
[ptb]
\begin{center}
\includegraphics[
height=1.0992in,
width=2.847in
]%
{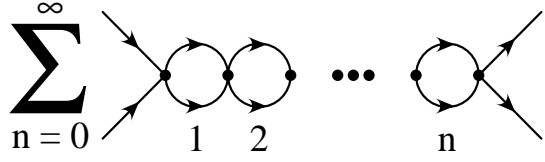}%
\caption{Sum over two-particle bubble diagrams.}%
\label{scattering}%
\end{center}
\end{figure}

We can calculate $C$ using a lattice regulator for various lattice spacings,
which denote as $a_{lattice}$. \ For simplicity we take the temporal lattice
spacing to be zero. \ We must sum all two-particle scattering bubble diagrams,
as shown in Fig. \ref{scattering}, and locate the pole in the scattering
amplitude.
We then use L\"{u}scher's formula for energy levels in a finite periodic box
\cite{Luscher:1986pf,Beane:2003da,Lee:2004qd} and tune the coefficients to
give the physically measured scattering lengths. \ From L\"{u}scher's formula
there should be a pole in the two-particle scattering amplitude with energy%
\begin{equation}
E_{pole}=\frac{4\pi a_{scatt}}{m_{N}L^{3}}+\cdots\text{,}%
\end{equation}
where $a_{scatt}$ is the scattering length. \ We can write the sum over bubble
diagrams as a geometric series. \ In order to produce a pole at this energy we
must have%

\begin{equation}
\frac{1}{m_{N}C}=\frac{1}{4\pi a_{scatt}}-\lim_{L\rightarrow\infty}\frac
{1}{a_{lattice}L^{3}}\sum_{\vec{k}\neq0}\frac{1}{6-2\cos\frac{2\pi k_{1}}%
{L}-2\cos\frac{2\pi k_{2}}{L}-2\cos\frac{2\pi k_{3}}{L}}, \label{pole}%
\end{equation}
where $a_{lattice}$ is the lattice spacing, and the sum is over integer values
$k_{1},k_{2},k_{3}$ from $0$ to $L-1$. \ Solving for $C$ gives%
\begin{equation}
C\simeq\frac{1}{m_{N}\left(  \frac{1}{4\pi a_{scatt}}-\frac{0.253}%
{a_{lattice}}\right)  }.
\end{equation}

For any chosen temperature and neutron density there is a corresponding
maximum value for the lattice spacing, $a_{lattice}.$ \ The requirements are
that the kinetic energy for the highest momentum mode must exceed the
temperature, and the lattice spacing must be less than the interparticle
spacing. \ We therefore have
\begin{equation}
a_{lattice}^{-1}\gg(a_{lattice}^{-1})_{\min}=\max\left[  \pi^{-1}\sqrt
{2m_{N}T},\rho^{1/3}\right]  .
\end{equation}
This sets an upper bound for the absolute value for the scale-dependent
coupling $C$,%
\begin{equation}
\left\vert C\right\vert \ll\left\vert C\right\vert _{\max}\equiv\frac{1}%
{m_{N}\left\vert \frac{1}{4\pi a_{scatt}}-0.253(a_{lattice}^{-1})_{\min
}\right\vert }\text{.}\label{Cmax}%
\end{equation}
This result will be useful for the shifted-field inequalities derived later.

\subsection{Convexity inequality}

The grand canonical partition function is given by%
\begin{equation}
Z_{G}(\mu,\mu_{3})=\int DnD\bar{n}\exp\left(  -S_{E}\right)  =\int DnD\bar
{n}\exp\left(  \int d^{4}x\,\mathcal{L}_{E}\right)  ,
\end{equation}
where we use the expression (\ref{first neutron}) for $\mathcal{L}_{E}$,%

\begin{equation}
\mathcal{L}_{E}=-\bar{n}[\partial_{4}-\tfrac{\vec{\nabla}^{2}}{2m_{N}}%
+(m_{N}^{0}-\mu-\mu_{3}\sigma_{3})]n-\tfrac{1}{2}C\bar{n}n\bar{n}n.
\end{equation}
Using a Hubbard-Stratonovich transformation
\cite{Stratonovich:1958,Hubbard:1959ub},\ we can rewrite $Z_{G}$ as%
\begin{equation}
Z_{G}\propto\int DnD\bar{n}Df\exp\left(  \int d^{4}x\,\mathcal{L}_{E}%
^{f}\right)  ,
\end{equation}
where%
\begin{equation}
\mathcal{L}_{E}^{f}=-\bar{n}[\partial_{4}-\tfrac{\vec{\nabla}^{2}}{2m_{N}%
}+(m_{N}^{0}-\mu-\mu_{3}\sigma_{3})]n+Cf\bar{n}n+\tfrac{1}{2}Cf^{2}.
\end{equation}

Let us define $\mathbf{M}$ as the matrix for the part of $\mathcal{L}_{E}^{f}$
bilinear in the neutron field,%
\begin{equation}
\mathbf{M}=-\left[  \partial_{4}-\tfrac{\vec{\nabla}^{2}}{2m_{N}}+(m_{N}%
^{0}-\mu-\mu_{3}\sigma_{3})\right]  +Cf.
\end{equation}
We observe that $\mathbf{M}$ has the block diagonal form,%
\begin{equation}
\mathbf{M}=\left[
\begin{array}
[c]{cc}%
M(\mu+\mu_{3}) & 0\\
0 & M(\mu-\mu_{3})
\end{array}
\right]  ,
\end{equation}
where%
\begin{equation}
M(\mu)=-\left[  \partial_{4}-\tfrac{\vec{\nabla}^{2}}{2m_{N}}+(m_{N}^{0}%
-\mu)\right]  +Cf\text{.}%
\end{equation}
Since $M$ is real valued, $\det M$ must also be real.

Integrating over the fermion fields gives us%
\begin{align}
Z_{G}(\mu,\mu_{3})  &  \propto\int DnD\bar{n}Df\exp\left(  \int d^{4}%
x\,\mathcal{L}_{E}^{f}\right) \nonumber\\
&  =\int D\Theta\det\mathbf{M=}\int D\Theta\det M(\mu+\mu_{3})\det M(\mu
-\mu_{3}),
\end{align}
where $D\Theta$ is the positive measure%
\begin{equation}
D\Theta=Df\exp\left(  \tfrac{1}{2}C\int d^{4}x\,f^{2}\right)  .
\end{equation}
\ Using the Cauchy-Schwarz inequality we find%
\begin{align}
\left\vert \int D\Theta\det M(\mu+\mu_{3})\det M(\mu-\mu_{3})\right\vert  &
\leq\int D\Theta\left\vert \det M(\mu+\mu_{3})\det M(\mu-\mu_{3})\right\vert
\nonumber\\
&  \leq\sqrt{\int D\Theta\left[  \det M(\mu+\mu_{3})\right]  ^{2}}\sqrt{\int
D\Theta\left[  \det M(\mu-\mu_{3})\right]  ^{2}}.
\end{align}
We can now compare the asymmetric partition function to the symmetric
partition function at chemical potentials $\mu+\mu_{3}$ and $\mu-\mu_{3}$,%
\begin{equation}
Z_{G}(\mu,\mu_{3})\leq\sqrt{Z_{G}(\mu+\mu_{3},0)}\sqrt{Z_{G}(\mu-\mu_{3}%
,0)}\text{.}%
\end{equation}

We now use the thermodynamic relation,%
\begin{equation}
\ln Z_{G}=\tfrac{PV}{k_{B}T}, \label{thermodynamic}%
\end{equation}
where $P$ is the pressure, $V$ is the volume, and $T$ is the temperature. \ We
find the upper bound%
\begin{equation}
P(\mu,\mu_{3})\leq\frac{1}{2}\left[  P(\mu+\mu_{3},0)+P(\mu-\mu_{3},0)\right]
.
\end{equation}

\subsection{Shifted-field inequality}

We start again with the grand canonical partition function%
\begin{equation}
Z_{G}(\mu,\mu_{3})=\int DnD\bar{n}\exp\left(  -S_{E}\right)  =\int DnD\bar
{n}\exp\left(  \int d^{4}x\,\mathcal{L}_{E}\right)  .
\end{equation}
This time we use the other expression (\ref{second neutron}) for
$\mathcal{L}_{E}$,%

\begin{equation}
\mathcal{L}_{E}=-\bar{n}[\partial_{4}-\tfrac{\vec{\nabla}^{2}}{2m_{N}}%
+(m_{N}^{0\prime}-\mu-\mu_{3}\sigma_{3})]n-\tfrac{1}{2}C^{\prime}\bar{n}%
\vec{\sigma}n\cdot\bar{n}\vec{\sigma}n.
\end{equation}
We can rewrite the grand canonical partition function using three
Hubbard-Stratonovich fields,
\begin{equation}
Z_{G}\propto\int DnD\bar{n}D\vec{\phi}\exp\left(  \int d^{4}x\,\mathcal{L}%
_{E}^{\vec{\phi}}\right)  ,
\end{equation}
where%

\begin{equation}
\mathcal{L}_{E}^{\vec{\phi}}=-\bar{n}[\partial_{4}-\tfrac{\vec{\nabla}^{2}%
}{2m_{N}}+(m_{N}^{0\prime}-\mu-\mu_{3}\sigma_{3})]n+iC^{\prime}\vec{\phi}%
\cdot\bar{n}\vec{\sigma}n-\tfrac{1}{2}C^{\prime}\vec{\phi}\cdot\vec{\phi}.
\end{equation}
Let $\mathbf{M}_{0}$ be the neutron matrix without the $\mu_{3}\sigma_{3}$
term$,$%
\begin{equation}
\mathbf{M}_{0}=-\left[  \partial_{4}-\tfrac{\vec{\nabla}^{2}}{2m_{N}}%
+(m_{N}^{0\prime}-\mu)\right]  +iC^{\prime}\vec{\phi}\cdot\vec{\sigma}.
\end{equation}
We note that%
\begin{equation}
\sigma_{2}\mathbf{M}_{0}\sigma_{2}=\mathbf{M}_{0}^{\ast},
\end{equation}
where $\mathbf{M}_{0}^{\ast}$ is the complex conjugate of $\mathbf{M}_{0}$.
\ This means that $\mathbf{M}_{0}$ is either singular, in which case
$\det\mathbf{M}_{0}=0$, or has the same eigenvalues as $\mathbf{M}_{0}^{\ast}%
$. \ In all cases $\det\mathbf{M}_{0}$ is real. \ Furthermore the fact that
$\sigma_{2}$ is antisymmetric means that the real eigenvalues of
$\mathbf{M}_{0}$ are doubly degenerate, and so $\det\mathbf{M}_{0}\geq0$
\cite{Hands:2000ei}.

We now concentrate on the part of $\mathcal{L}_{E}^{\vec{\phi}}$ that contains
$\mu_{3}$ and $\phi_{3}$,%
\begin{equation}
-\tfrac{1}{2}C^{\prime}\phi_{3}^{2}+iC^{\prime}\phi_{3}\bar{n}\sigma_{3}%
n+\mu_{3}\bar{n}\sigma_{3}n.
\end{equation}
We can rewrite this as%
\begin{equation}
-\tfrac{1}{2}C^{\prime}\phi_{3}^{\prime2}-i\mu_{3}\phi_{3}^{\prime}%
+iC^{\prime}\phi_{3}^{\prime}\bar{n}\sigma_{3}n+\tfrac{1}{2}\tfrac{\mu_{3}%
^{2}}{C^{\prime}}%
\end{equation}
where%
\begin{equation}
\phi_{3}^{\prime}=\phi_{3}-i\tfrac{\mu_{3}}{C^{\prime}}.
\end{equation}
The original contour of integration for $\phi_{3}^{\prime}$ is off the real
axis, but we can deform the contour onto the real axis. \ For notational
convenience we now drop the prime on $\phi_{3}^{\prime}$ and have%
\begin{equation}
\mathcal{L}_{E}^{\vec{\phi}}=-\bar{n}[\partial_{4}-\tfrac{\vec{\nabla}^{2}%
}{2m_{N}}+(m_{N}^{0\prime}-\mu)]n+iC^{\prime}\vec{\phi}\cdot\bar{n}\vec
{\sigma}n-\tfrac{1}{2}C^{\prime}\vec{\phi}\cdot\vec{\phi}-i\mu_{3}\phi
_{3}+\tfrac{1}{2}\tfrac{\mu_{3}^{2}}{C^{\prime}}.
\end{equation}
The neutron matrix is now $\mathbf{M}_{0}$, which we have shown has a
non-negative determinant. \ The complex phase is contained entirely in the
local expression $-i\mu_{3}\phi_{3}$.

We now have%
\begin{align}
Z_{G}  &  \propto\int D\Theta\exp\left\{  \int d^{4}x\left[  -i\mu_{3}\phi
_{3}+\tfrac{1}{2}\tfrac{\mu_{3}^{2}}{C^{\prime}}\right]  \right\} \nonumber\\
&  =\exp(\tfrac{V\mu_{3}^{2}}{2C^{\prime}k_{B}T})\int D\Theta\exp\left(
-i\mu_{3}\int d^{4}x\;\phi_{3}\right)  ,
\end{align}
where $D\Theta$ is the normalized positive measure
\begin{equation}
D\Theta=\frac{D\vec{\phi}\;\det\mathbf{M}_{0}\exp\left(  -\int d^{4}%
x\,\mathcal{V}(\vec{\phi})\right)  }{\int D\vec{\phi}\;\det\mathbf{M}_{0}%
\exp\left(  -\int d^{4}x\,\mathcal{V}(\vec{\phi})\right)  }%
\end{equation}
with
\begin{equation}
-\,\mathcal{V}(\vec{\phi})=-\tfrac{1}{2}C^{\prime}\vec{\phi}\cdot\vec{\phi}.
\end{equation}
\qquad

Using (\ref{thermodynamic}) we find%
\begin{align}
P(\mu,\mu_{3})-P(\mu,0) &  =\tfrac{k_{B}T}{V}\ln\left[  \exp(\tfrac{V\mu
_{3}^{2}}{2C^{\prime}k_{B}T})\int D\Theta\exp\left(  -i\mu_{3}\int
d^{4}x\;\phi_{3}\right)  \right]  \nonumber\\
&  =\tfrac{\mu_{3}^{2}}{2C^{\prime}}+\tfrac{k_{B}T}{V}\ln\left[  \int
D\Theta\exp\left(  -i\mu_{3}\int d^{4}x\;\phi_{3}\right)  \right]  .
\end{align}
So we conclude that%
\begin{equation}
P(\mu,\mu_{3})\leq P(\mu,0)+\tfrac{\mu_{3}^{2}}{2C^{\prime}}%
.\label{cprime neutron}%
\end{equation}
This upper bound is unusual in that it relates physical observables
independent of the cutoff scale to the scale-dependent coupling $C^{\prime}$.
\ By taking the lattice spacing as large as possible, we have
\begin{equation}
C^{\prime}=\tfrac{1}{3}\left\vert C\right\vert _{\max},
\end{equation}
where $\left\vert C\right\vert _{\max}$ was defined in (\ref{Cmax}), and
therefore%
\begin{equation}
P(\mu,\mu_{3})\leq P(\mu,0)+\tfrac{3\mu_{3}^{2}}{2\left\vert C\right\vert
_{\max}}.\label{shifted neutron}%
\end{equation}
\ As a rough estimate of the quantities involved, we note that for $\rho
\sim0.1\rho_{N}$ and $T<10$ MeV, $\left\vert C\right\vert _{\max}$ is about
$3$ fm$^{2}$.

As $C^{\prime}$ decreases the upper bound in (\ref{cprime neutron}) increases.
\ But at the same time the tightness of the bound becomes poorer as complex
phase oscillations due to the term%
\begin{equation}
\exp\left[  \int d^{4}x\left(  -\tfrac{1}{2}C^{\prime}\phi_{3}^{2}-i\mu
_{3}\phi_{3}\right)  \right]
\end{equation}
become more significant. \ The average phase for our functional integral is
given by%
\begin{align}
\left\langle \text{phase}\right\rangle  &  =\int D\Theta\exp\left(  -i\mu
_{3}\int d^{4}x\;\phi_{3}\right) \nonumber\label{local}\\
&  =\exp\left[  \tfrac{V}{k_{B}T}\left(  P(\mu,\mu_{3})-P(\mu,0)-\tfrac
{\mu_{3}^{2}}{2C^{\prime}}\right)  \right]  .
\end{align}

Given an estimate of the pressure difference, this relation can be used to
predict the feasibility of a numerical simulation using this representation of
the functional integral. \ In cases where the phase problem is not too severe
we can use hybrid Monte Carlo to generate Hubbard-Stratonovich field
configurations according to the relative probability weight $\det
\mathbf{M}_{0}$. \ The phase of the configuration can then be included as an
observable using the local expression $-i\mu_{3}\phi_{3}$. \ This local
expression for the phase could increase algorithmic speed by several orders of
magnitude. \ The only known way to compute the phase of matrix determinants is
LU decomposition, an algorithm which writes a matrix as a product of lower and
upper triangular matrices. \ The number of operations for LU decomposition
scales as $N^{3}$, where $N$ is the dimension of the matrix. \ For an $L^{4}$
lattice the scaling is thus $L^{12}$.

\section{Four fermion states - $SU(2)\times SU(2)$}

We now consider an effective theory with four species of interacting fermions
and an $SU(2)\times SU(2)$ symmetry. \ Let $N$ be a quartet of fermion states,
which we can regard as nucleon fields,%
\begin{equation}
N=\left[
\begin{array}
[c]{c}%
p\\
n
\end{array}
\right]  \otimes\left[
\begin{array}
[c]{c}%
\uparrow\\
\downarrow
\end{array}
\right]  .
\end{equation}
We use $p$($n$) to represent protons(neutrons) and $\uparrow$($\downarrow$) to
represent up(down) spins. \ We use $\vec{\tau}$ to represent Pauli matrices
acting in isospin space and $\vec{\sigma}$ to represent Pauli matrices acting
in spin space. \ We assume exact isospin and spin symmetry in the absence of
symmetry-breaking chemical potentials, and so the symmetry group is
$SU(2)_{I}\times SU(2)_{S}$.

In the non-relativistic limit and below the threshold for pion production, we
can write the lowest-order terms in the effective Lagrangian in two equivalent
ways,%
\begin{align}
\mathcal{L}_{E}  &  =-\bar{N}[\partial_{4}-\tfrac{\vec{\nabla}^{2}}{2m_{N}%
}+(m_{N}^{0}-\mu)]N-\tfrac{1}{2}C_{S}(\bar{N}N)^{2}-\tfrac{1}{2}C_{T}\bar
{N}\vec{\sigma}N\cdot\bar{N}\vec{\sigma}N\nonumber\\
&  -\tfrac{1}{3!}C_{3}(\bar{N}N)^{3}-\tfrac{1}{4!}C_{4}(\bar{N}N)^{4},
\label{C_T}%
\end{align}
or%
\begin{align}
\mathcal{L}_{E}  &  =-\bar{N}[\partial_{4}-\tfrac{\vec{\nabla}^{2}}{2m_{N}%
}+(m_{N}^{0\prime}-\mu)]N-\tfrac{1}{2}C_{S}^{\prime}(\bar{N}N)^{2}-\tfrac
{1}{2}C_{U}^{\prime}\bar{N}\vec{\tau}N\cdot\bar{N}\vec{\tau}N\nonumber\\
&  -\tfrac{1}{3!}C_{3}(\bar{N}N)^{3}-\tfrac{1}{4!}C_{4}(\bar{N}N)^{4}.
\label{C_U}%
\end{align}
We will introduce symmetry breaking chemical potentials later. \ We have
included both three-body and four-body forces. \ The $SU(4)$-symmetric
three-nucleon force is needed for consistent renormalization and has been
shown to be the dominant three-body force contribution
\cite{Mehen:1999qs,Bedaque:1998kg,Bedaque:1999ve}.

With four distinct fermion species there are two irreducible representations
of $SU(2)_{I}\times SU(2)_{S}$ for two fermions in an s-wave, a spin-singlet
isospin-triplet $(S=0)$ or an isospin-singlet spin-triplet $(I=0)$. \ One can
show that \cite{Lee:2004ze}%
\begin{equation}
C_{U}^{\prime}=-C_{T},\quad C_{S}^{\prime}=C_{S}-2C_{T}.
\end{equation}
In the case of nucleons, one finds that both of the s-wave channels are
attractive, with the $I=0$ channel being more strongly attractive,%
\begin{equation}
\frac{1}{a_{scatt}^{I=0}}>\frac{1}{a_{scatt}^{S=0}}.
\end{equation}
This implies that \cite{Lee:2004ze}%
\begin{align}
C_{S}  &  <3C_{T},\quad C_{T}<0,\\
C_{S}^{\prime}  &  <-C_{U}^{\prime},\quad C_{U}^{\prime}>0.
\end{align}
For a more general system with four fermion states and an $SU(2)\times SU(2)$
symmetry, we can interchange the isospin and spin labels so that, without loss
of generality,%
\begin{equation}
\frac{1}{a_{scatt}^{I=0}}\geq\frac{1}{a_{scatt}^{S=0}}.
\end{equation}
In the special case when the scattering lengths are equal, the symmetry group
is the full Wigner $SU(4)$ symmetry \cite{Wigner:1939a}, and the isospin and
spin labels can be interchanged.

\subsection{Two-body operator coefficients}

We determine the two-body operator coefficients in the same manner as before.
\ The only difference is that there are now two s-wave channels. \ The
coefficient $C$ in (\ref{pole}) is replaced by $C^{S=0}$ and $C^{I=0}$ where%
\begin{align}
C^{S=0} &  =C_{S}^{\prime}+C_{U}^{\prime},\\
C^{I=0} &  =C_{S}^{\prime}-3C_{U}^{\prime}\text{.}%
\end{align}
We then find%
\begin{align}
C_{S}^{\prime} &  \simeq\frac{3}{4m_{N}\left(  \frac{1}{4\pi a_{scatt}^{S=0}%
}-\frac{0.253}{a_{lattice}}\right)  }+\frac{1}{4m_{N}\left(  \frac{1}{4\pi
a_{scatt}^{I=0}}-\frac{0.253}{a_{lattice}}\right)  },\\
C_{U}^{\prime} &  \simeq\frac{1}{4m_{N}\left(  \frac{1}{4\pi a_{scatt}^{S=0}%
}-\frac{0.253}{a_{lattice}}\right)  }-\frac{1}{4m_{N}\left(  \frac{1}{4\pi
a_{scatt}^{I=0}}-\frac{0.253}{a_{lattice}}\right)  }\text{.}%
\end{align}
For any chosen temperature and nucleon density there is again a corresponding
maximum value for the lattice spacing,%
\begin{equation}
a_{lattice}^{-1}\gg(a_{lattice}^{-1})_{\min}=\max\left[  \pi^{-1}\sqrt
{2m_{N}T},\rho^{1/3}\right]  .
\end{equation}
This sets a maximum value for the absolute value of the coupling
$C_{U}^{\prime}$,%
\begin{equation}
\left\vert C_{U}^{\prime}\right\vert \ll\left\vert C_{U}^{\prime}\right\vert
_{\max}\equiv\frac{\left\vert \frac{1}{4\pi a_{scatt}^{I=0}}-\frac{1}{4\pi
a_{scatt}^{S=0}}\right\vert }{4m_{N}\left\vert \left(  \frac{1}{4\pi
a_{scatt}^{S=0}}-0.253(a_{lattice}^{-1})_{\min}\right)  \left(  \frac{1}{4\pi
a_{scatt}^{I=0}}-0.253(a_{lattice}^{-1})_{\min}\right)  \right\vert
}.\label{CUmax}%
\end{equation}
A similar bound for $C_{S}^{\prime}$ can be made but is not needed in our analysis.

\subsection{Convexity inequality for $\mu_{3}^{S}$}

We first consider the case when an asymmetric chemical potential $\mu_{3}^{S}$
is coupled to the nucleon spins. \ The grand canonical partition function is
given by%
\begin{equation}
Z_{G}=\int DND\bar{N}\exp\left(  -S_{E}\right)  =\int DND\bar{N}\exp\left(
\int d^{4}x\,\mathcal{L}_{E}\right)  ,
\end{equation}
where we take the form of $\mathcal{L}_{E}$ given in (\ref{C_U}) with an
asymmetric spin chemical potential,%
\begin{align}
\mathcal{L}_{E}  &  =-\bar{N}[\partial_{4}-\tfrac{\vec{\nabla}^{2}}{2m_{N}%
}+(m_{N}^{0\prime}-\mu-\mu_{3}^{S}\sigma_{3})]N-\tfrac{1}{2}C_{S}^{\prime
}(\bar{N}N)^{2}-\tfrac{1}{2}C_{U}^{\prime}\bar{N}\vec{\tau}N\cdot\bar{N}%
\vec{\tau}N\nonumber\\
&  -\tfrac{1}{3!}C_{3}(\bar{N}N)^{3}-\tfrac{1}{4!}C_{4}(\bar{N}N)^{4}.
\end{align}
Using Hubbard-Stratonovich transformations we can rewrite $Z_{G}$ as%
\begin{equation}
Z_{G}\propto\int DND\bar{N}DfD\vec{\phi}\exp\left(  \int d^{4}x\,\mathcal{L}%
_{E}^{f,\vec{\phi}}\right)  ,
\end{equation}
where%
\begin{align}
\mathcal{L}_{E}^{f,\vec{\phi}}  &  =-\bar{N}[\partial_{4}-\tfrac{\vec{\nabla
}^{2}}{2m_{N}}+(m_{N}^{0\prime}-\mu-\mu_{3}^{S}\sigma_{3})]N+f\bar{N}%
N+iC_{U}^{\prime}\vec{\phi}\cdot\bar{N}\vec{\tau}N\nonumber\\
&  +g(f)-\tfrac{1}{2}C_{U}^{\prime}\vec{\phi}\cdot\vec{\phi}.
\end{align}
In \cite{Chen:2004rq} it was shown that three-body and four-body forces can be
introduced without spoiling positivity of the functional integral measure.
\ The only requirements are that the three-body force is not too strong and
the four-body force is not too repulsive. \ Estimates of the three- and
four-body forces suggest that these conditions are satisfied. \ For our
analysis here we assume that to be the case, and the function $g(f)$ is a
real-valued function which produces the two-, three-, and four-body force
terms involving $\bar{N}N$.

The nucleon matrix $\mathbf{M}$ has the block diagonal structure%
\begin{equation}
\mathbf{M}=\left[
\begin{array}
[c]{cc}%
M(\mu+\mu_{3}^{S}) & 0\\
0 & M(\mu-\mu_{3}^{S})
\end{array}
\right]  ,
\end{equation}
where the upper block is for up spins and the lower block is for down spins.
$\ M$ is a matrix in isospin space,%

\begin{equation}
M(\mu)=-\left[  \partial_{4}-\tfrac{\vec{\nabla}^{2}}{2m_{N}}+(m_{N}^{0\prime
}-\mu)\right]  +f+iC_{U}^{\prime}\vec{\phi}\cdot\vec{\tau}.
\end{equation}
We note that
\begin{equation}
\tau_{2}M\tau_{2}=M^{\ast},
\end{equation}
and so $\det M\geq0$.

Integrating over the fermion fields gives us%
\begin{align}
Z_{G}(\mu,\mu_{3}^{S})  &  \propto\int DND\bar{N}DfD\vec{\phi}\exp\left(  \int
d^{4}x\,\mathcal{L}_{E}^{f,\vec{\phi}}\right) \nonumber\\
&  =\int D\Theta\det\mathbf{M=}\int D\Theta\det M(\mu+\mu_{3}^{S})\det
M(\mu-\mu_{3}^{S}),
\end{align}
where%
\begin{equation}
D\Theta=DfD\vec{\phi}\exp\left(  -\int d^{4}x\,\mathcal{V}(f,\vec{\phi
})\right)
\end{equation}
with
\begin{equation}
-\,\mathcal{V}(f,\vec{\phi})=g(f)-\tfrac{1}{2}C_{U}^{\prime}\vec{\phi}%
\cdot\vec{\phi}.
\end{equation}
From the Cauchy-Schwarz inequality we get%
\begin{equation}
Z_{G}(\mu,\mu_{3})\leq\sqrt{Z_{G}(\mu+\mu_{3}^{S},0)}\sqrt{Z_{G}(\mu-\mu
_{3}^{S},0)}\text{.}%
\end{equation}
We therefore find an upper bound for the pressure,%
\begin{equation}
P(\mu,\mu_{3}^{S})\leq\frac{1}{2}\left[  P(\mu+\mu_{3}^{S},0)+P(\mu-\mu
_{3}^{S},0)\right]  .
\end{equation}

\subsection{Shifted-field inequality for $\mu_{3}^{I}$}

We now consider the case with an isospin chemical potential $\mu_{3}^{I}$.
\ We start with the Lagrange density in terms of the Hubbard-Stratonovich
fields,
\begin{align}
\mathcal{L}_{E}^{f,\vec{\phi}}  &  =-\bar{N}[\partial_{4}-\tfrac{\vec{\nabla
}^{2}}{2m_{N}}+(m_{N}^{0\prime}-\mu-\mu_{3}^{I}\tau_{3})]N+f\bar{N}%
N+iC_{U}^{\prime}\vec{\phi}\cdot\bar{N}\vec{\tau}N\nonumber\\
&  +g(f)-\tfrac{1}{2}C_{U}^{\prime}\vec{\phi}\cdot\vec{\phi}.
\end{align}
Let $\mathbf{M}_{0}$ be the nucleon matrix without the $\mu_{3}^{I}\tau_{3}$
term$,$%
\begin{equation}
\mathbf{M}_{0}=-\left[  \partial_{4}-\tfrac{\vec{\nabla}^{2}}{2m_{N}}%
+(m_{N}^{0\prime}-\mu)\right]  +f+iC_{U}^{\prime}\vec{\phi}\cdot\vec{\tau}.
\end{equation}
We note that%
\begin{equation}
\tau_{2}\mathbf{M}_{0}\tau_{2}=\mathbf{M}_{0}^{\ast}, \label{tau_2}%
\end{equation}
and so $\det\mathbf{M}_{0}\geq0.$

As we did for the two fermion case, we now shift the $\phi_{3}$ field and find
the inequality%
\begin{equation}
P(\mu,\mu_{3}^{I})\leq P(\mu,0)+\tfrac{(\mu_{3}^{I})^{2}}{2C_{U}^{\prime}%
}.\label{mainresult}%
\end{equation}
If we take the lattice spacing as large as possible then%
\begin{equation}
P(\mu,\mu_{3}^{I})\leq P(\mu,0)+\tfrac{(\mu_{3}^{I})^{2}}{2\left\vert
C_{U}^{\prime}\right\vert _{\max}},
\end{equation}
where $\left\vert C_{U}^{\prime}\right\vert _{\max}$ was defined in
(\ref{CUmax}). \ As a rough estimate of the quantities involved, we note that
for $\rho\sim0.1\rho_{N}$ and $T<10$ MeV, $\left\vert C_{U}^{\prime
}\right\vert _{\max}$ is about $0.2$ fm$^{2}$. \ In this case however the
situation is complicated by nuclear saturation, and it is not clear that the
pionless effective theory is applicable.

\section{Summary and discussion}

The main results we have shown are as follows. \ We first considered the two
fermion system with an attractive interaction and $SU(2)$ symmetry. \ If $\mu$
is the symmetric chemical potential and $\mu_{3}$ is the asymmetric chemical
potential, we proved both the convexity inequality%
\begin{equation}
P(\mu,0)\leq P(\mu,\mu_{3})\leq\frac{1}{2}\left[  P(\mu+\mu_{3},0)+P(\mu
-\mu_{3},0)\right]  , \label{1}%
\end{equation}
and the shift-field inequality%
\begin{equation}
P(\mu,0)\leq P(\mu,\mu_{3})\leq P(\mu,0)+\tfrac{3\mu_{3}^{2}}{2\left\vert
C\right\vert _{\max}}. \label{2}%
\end{equation}

We then analyzed the four fermion system with an $SU(2)_{I}\times SU(2)_{S}$
symmetry. \ We considered the case when both s-wave channels are attractive
and without loss of generality assumed the $I=0$ channel to be more strongly
attractive. \ With $\mu$ as the symmetric chemical potential and $\mu_{3}^{S}$
as the asymmetric spin chemical potential we proved the convexity inequality%
\begin{equation}
P(\mu,0)\leq P(\mu,\mu_{3}^{S})\leq\frac{1}{2}\left[  P(\mu+\mu_{3}%
^{S},0)+P(\mu-\mu_{3}^{S},0)\right]  . \label{3}%
\end{equation}
For non-zero asymmetric isospin chemical potential $\mu_{3}^{I}$ we proved the
shifted-field inequality%

\begin{equation}
P(\mu,0)\leq P(\mu,\mu_{3}^{I})\leq P(\mu,0)+\tfrac{(\mu_{3}^{I})^{2}%
}{2\left\vert C_{U}^{\prime}\right\vert _{\max}}. \label{4}%
\end{equation}
In the Wigner $SU(4)$ symmetry limit, we note that the shift-field inequality
(\ref{4}) becomes meaningless since $\left\vert C_{U}^{\prime}\right\vert
_{\max}\rightarrow0$. \ However in this limit we also have the convexity
inequality for $\mu_{3}^{I}$,%
\begin{equation}
P(\mu,0)\leq P(\mu,\mu_{3}^{I})\leq\frac{1}{2}\left[  P(\mu+\mu_{3}%
^{I},0)+P(\mu-\mu_{3}^{I},0)\right]  . \label{Wigner}%
\end{equation}

The equation of state for nuclear matter with small isospin asymmetries can be
measured indirectly in the laboratory by studying nuclear multifragmentation.
\ Of the inequalities presented here, the simplest and perhaps most
interesting to check is the isospin convexity inequality (\ref{Wigner}) in the
Wigner symmetry limit. \ Since much is still unknown about asymmetric nuclear
matter, this Wigner pressure inequality may be a useful consistency check for
proposed phenomenological models for asymmetric nuclear matter.

While some of the inequalities are difficult to observe in nuclear physics
experiments,\ each of our results could be tested in the cold Fermi gas system
where parameters in the effective Lagrangian can be tuned. \ Such experiments
can in principle test the inequalities over a range of physical parameters and
probe universal results in the limit of infinite scattering length and zero
range. \ Although four fermion systems have not yet been produced, these may
be possible in the near future.

On the computational side, the inequalities can also be checked by
non-perturbative lattice simulations. \ There have been several recent
simulations of effective theories on the lattice
\cite{Muller:1999cp,Lee:2004si,Wingate:2004wm,Lee:2004qd}. \ It will be
particularly interesting to look at symmetric and asymmetric nuclear matter in
the Wigner symmetry limit, which can be simulated without any sign problem.

It remains to be seen how well many-body nucleon systems can be described
without explicit pions. \ Results from \cite{Lee:2004qd} for dilute neutron
matter suggest that lowest-order effective field theory without pions works
very well in describing the neutron equation of state. \ The situation for
nearly symmetric nuclear matter, however, is less clear due to the effect of
saturation which requires higher densities.

With pions included the effective theory action can in general become
negative. \ This would in principle invalidate any inequality\ based on
positivity of the action. \ However it has been shown that this sign problem
goes away in the static limit \cite{Chandrasekharan:2003wy}. \ Furthermore the
sign problem has been numerically observed to be small \cite{Lee:2004si} in
simulations with neutrons and neutral pions for temperatures above 10 MeV and
densities at or below normal nuclear matter density. \ If one neglects these
sign changes, then the sign-quenched results for the effective theory with
pions will also satisfy each of the inequalities proven here\textit{.}

The author thanks Jiunn-Wei Chen and Thomas Schaefer for several helpful
disucssions. \ This work was supported by Department of Energy grant DE-FG02-04ER41335.

\textrm{
\bibliographystyle{apsrev}
\bibliography{NuclearMatter}
}

\end{document}